\documentclass[12pt]{article}
\usepackage{epsfig}
\usepackage{graphics}
\usepackage{amssymb,amsmath}
\usepackage{wasysym}
\usepackage{cite}
\usepackage[footnotesize,bf]{caption}
\usepackage{axodraw}
\usepackage{color}
\usepackage{verbatim}
\usepackage{slashed}
\usepackage{subfigure}
\usepackage[active]{srcltx} 
\usepackage{cite}
\usepackage{bbm}
\usepackage{stmaryrd}
\newcommand{\lsim}{
\mathrel{\hbox{\rlap{\hbox{\lower4pt\hbox{$\sim$}}}\hbox{$<$}}}}

\newcommand{\gsim}{
\mathrel{\hbox{\rlap{\hbox{\lower4pt\hbox{$\sim$}}}\hbox{$>$}}}}
\newcommand{\pmatr}[1] {\begin{pmatrix} #1 \end{pmatrix}}
\newcommand{\tev}{\, {\rm TeV}}
\newcommand{\gev}{\, {\rm GeV}}

\def\sla #1 {\slashed{#1}}

\def\simge{\mathrel{\rlap{\raise 0.511ex \hbox{$>$}}{\lower 0.511ex \hbox{$\sim$}}}}
\def\simle{\mathrel{\rlap{\raise 0.511ex \hbox{$<$}}{\lower 0.511ex \hbox{$\sim$}}}}
\def\slash#1{\setbox0=\hbox{$#1$}\dimen0=\wd0
      \setbox1=\hbox{/} \dimen1=\wd1 \ifdim\dimen0>\dimen1
      \rlap{\hbox to \dimen0{\hfil/\hfil}} #1                        \else
      \rlap{\hbox to \dimen1{\fil$#1$\hfil}}
      /   \fi}

\def\eps{\varepsilon}

\newcommand{\be}{\begin{equation}}
\newcommand{\ee}{\end{equation}}
\newcommand{\bea}{\begin{eqnarray}}
\newcommand{\eea}{\end{eqnarray}}

\newcommand{\bi}{\begin{itemize}}
\newcommand{\ei}{\end{itemize}}

\begin{document}
\begin{titlepage}
\vspace*{-0.5truecm}

\begin{flushright}
TUM-HEP-675/07\\
UAB-FT/637\\
\end{flushright}

\vspace*{0.3truecm}

\begin{center}
{\Large{\bf Bounding the Minimal 331 Model\\
\vspace*{0.3truecm}
through the Decay $B \to X_s \gamma$}}
\end{center}

\vspace{0.9truecm}

\begin{center}
{\bf Christoph Promberger${}^a$, 
Sebastian Schatt${}^a$, Felix Schwab${}^{b}$ and Selma Uhlig${}^a$}
 
\vspace{0.5truecm}

${}^a$ {\sl Physik Department, Technische Universit\"at M\"unchen,
D-85748 Garching, Germany}\\

\vspace{0.2truecm}

${}^b$ {\sl Departament de F\'{\i}sica Te\`orica, IFAE, UAB, E-08193 Bellaterra,
 Barcelona, Spain}

\end{center}

\vspace{0.6cm}
\begin{abstract}
\vspace{0.2cm}\noindent
We study the decay $B \to X_s \gamma$ within the framework of the minimal 331 model, taking into account both new experimental and theoretical developments that 
allow us to update and improve on an existing ten year old analysis. In contrast to several other flavor changing observables that are modified already at tree level 
from a new $Z'$ gauge boson, we have only one loop contributions in this case. Nevertheless, these are interesting, as they may be enhanced and can shed light on the charged gauge boson and Higgs sector of the model. Numerically, we find that the Higgs sector, which is well approximated by a 2 Higgs doublet
model (2HDM), dominates, since the gauge contributions are already very strongly constrained. With respect to $B \to X_s \gamma$, the signal of the minimal 331 model is therefore nearly identical to the 2HDM one, which allows us to obtain a 
lower bound on the charged Higgs mass. Further, we observe, in analogy to the 2HDM model, that the branching fraction can be rather strongly increased for small values of 
$\tan \beta$. Also, we find that $B \to X_s \gamma$ has no impact on the bounds obtained on rare $K$ and $B$ decays in an earlier analysis.
\end{abstract}

\vspace*{0.5truecm}
\vfill
\noindent

%
%
%
\end{titlepage}
%
%
%



%
%
%
\section{Introduction}\label{sec:intro}
Presently, all phenomena observed in nature (with the exception of gravity) are described within the standard model (SM) of particle physics. This model seems
to work beautifully up to scales of at least $\mathcal{O}(\tev)$. On the other hand, successful as the SM may be, the general hope (and belief) of particle 
physicists is that it should be a part of a more fundamental theory. This belief is based on several theoretical shortcomings of the model, the most important of
which are the instability of the Higgs mass as well as the general particle content of the model, such as the fact that each fermion
type appears in three generations. This last point is addressed in the context of 331 models \cite{Frampton:1992wt,Pisano:1991ee}, where the requirement of 
anomaly cancellation combined with the asymptotic freedom of QCD force the number of generations to be exactly three. To achieve this, the electroweak $SU(2)_L$ 
of the SM is extended to an $SU(3)_L$, where the third generation is treated differently, i.e. is transformed as an anti-triplet. 

This set-up leads to the existence of several new particles, in particular
of new gauge bosons, such as a $Z'$ that transmits flavor changes at tree level due to the different treatment of the third generation. As a consequence, this
gives rise to tree level contributions for several observables that proceed at loop level only in the SM and which have been extensively studied in the literature
\cite{Liu:1993gy,GomezDumm:1994tz,Rodriguez:2004mw,PSS,Martinez:2008jj}.
In addition to these observables, it is also interesting to investigate the inclusive decay $B \to X_s \gamma $, 
where $X_s$ denotes a sum over all final states containing a strange quark. While tree level contributions are absent here, those at
one loop may be interesting, and have first been studied for the minimal model (to which we will also restrict ourselves) in \cite{Agrawal:1995vp}, 
after similar, analogous effects to $\varepsilon'/\varepsilon$ have been 
investigated in \cite{Agrawal:1994pf}. There are several reasons that make this analysis worthwhile, in spite of the fact that a first reasoning would suggest to focus on
tree level $Z'$ exchange only. These are:
\begin{itemize}
\item The mechanism that causes FCNCs at tree level (i.e. the different treatment of the third generation) also enhances the one loop contribution rather 
strongly. This is due to the breakdown of the GIM mechanism, which acts very effectively in the SM. 
In any case, it was found in \cite{Agrawal:1995vp} that the new contributions could, in principle, be almost comparable in size to  
the SM. Since the one loop effects are also governed by other particles than the tree level ones, i.e. charged gauge bosons and Higgs fields, 
one may also hope to shed some light on an entirely different sector of the model. 
\item In the ten years since the analysis of \cite{Agrawal:1995vp}, the data on $B \to X_s \gamma$ have improved considerably 
\cite{Koppenburg:2004fz,Aubert:2005cua,Aubert:2006gg}, leading to a
 very precise experimental number.
This can be used to place constraints on the parameter space of the model
which can be compared with those coming from other FCNC processes, obtained for example in \cite{PSS}.
\end{itemize}
In view of this, we update the $B \to X_s \gamma$ analysis of \cite{Agrawal:1995vp} using new data, retaining the general scheme of that analysis. This concerns in 
particular the treatment of the QCD corrections via the renormalization group equations, and the treatment of the Higgs sector. Here, we study only an effective 2 Higgs 
doublet model 
(2HDM) type II, which is a good approximation in the flavor sector \cite{Liu:1993gy,Agrawal:1995vp}. This analysis extends and concludes our study of FCNC processes 
begun in \cite{PSS}. Our paper is then organized as follows: First, in Section \ref{sec:model}, we introduce very briefly the minimal 331 model, focusing on those parts of 
the model that are important for the penguin diagrams contributing. Next, we give the most general background on $B \to X_s \gamma$ in Section
\ref{sec:obs}, where we also list and explain the different contributions to the decay and discuss the cancellation of divergences. 
Section \ref{sec:numerics} then contains all our numerics, where we compare these new
constraints with those from the measurement of the $B_s^0$ mass difference $\Delta M_s$. These are the only relevant ones in this context, since the 
different sectors of flavor transitions $sd$, $bd$ and $bs$ decouple. Finally, we conclude in Section \ref{sec:conclusions}.

\section{The Minimal 331 Model}\label{sec:model}
The minimal 331 model has been discussed extensively in
  \cite{Liu:1993gy,Ng:1992st,Liu:1994rx} and in short in \cite{PSS}, from where we take the conventions.
Many variations of this minimal model have been developed, such as
ones with right handed neutrinos, supersymmetric versions, and others
\cite{Foot:1994ym,Montero:1992jk,Long:1999ij,Gutierrez:2004sb,Diaz:2004fs,Okamoto:1999cf,Kitabayashi:2000nq,Tully:2000kk,Montero:2001ts,Cortez:2005cp,Duong:1993zn,Montero:2000ng,Montero:2004uy,Ponce:2002sg}.
Generally, in 331 models the electroweak $SU(2)_L$ of the SM is extended to a $SU(3)_L$, which is broken down in two steps: 
\begin{equation}\label{Breaking}
SU(3)_C \times SU(3)_L \times U(1)_X \stackrel{v_{\sigma}}{\Rightarrow} SU(3)_C \times SU(2)_L \times U(1)_Y
\stackrel{v_{\eta},v_{\rho}}{\Rightarrow} SU(3)_C \times U(1)_{em}
\end{equation}
Evidently, this breaking process involves three Higgs doublets, one of which develops a vacuum expectation value (VEV) at a high scale, while the other two VEVs are of  the order of the weak scale. 
This leads to a much richer scalar sector than in the SM, which has been extensively studied for the minimal model in \cite{Anh:2000bs,Diaz:2003dk}, as well as a more 
complicated Yukawa structure \cite{Liu:1993gy,Ng:1992st,Liu:1994rx,GomezDumm:1994tz,Diaz:2004fs}. Physically, after absorbing the appropriate number of Goldstone bosons,
one is left with one light neutral Higgs scalar, 7 heavier neutral Higgs particles as well as 4 singly and 3 doubly charged Higgs 
particles, all of which are heavy. On the other hand, the quark
doublets of the SM are extended to triplets by adding an additional
heavy quark. All of the quarks, as well as their right-handed
counterparts, have three quantum numbers called $X$, $T_3$ and $T_8$, corresponding to the diagonal generators of $SU(3)$.  

From these, the electric charge can be obtained by 
\begin{equation} \label{charge}
Q=T_3+\beta\, T_8 + X; \qquad \beta= \sqrt 3
\end{equation}
in our normalization of the hypercharge $X$. To ensure anomaly cancellation, the third generation couples differently than the first two. This has the important consequence
of generating flavor changing neutral currents at tree level. 
Another
curiosity of the minimal 331 model is the Landau Pole that can arise at rather low energies. It becomes apparent when one expresses the ratio of $SU(3)$ and $SU(2)$ couplings through the Weinberg angle as
\begin{equation} \label{WW}
\frac{g_X^2}{g^2} = \frac{6 \, \sin^2 \theta_W}{1 \!-\!4 \: \sin^2 \theta_W} \,.
\end{equation} 
Clearly, the theory is ill defined if $\sin^2 \theta_W$ grows to be 1/4, which puts an upper limit on the scale of the symmetry breaking as well as on the heavy gauge boson 
masses. Analyzing this carefully \cite{LP}, gives an upper bound of several $\tev$ for the $Z'$ mass, so that in the numerical analysis below we will follow \cite{PSS} and 
use values of up to $M_{Z'}=5 \tev$. 
In choosing $\beta=\sqrt 3$ in (\ref{charge}), and assigning the
appropriate hypercharge structure, we have distinguished the model to
be the {\em minimal} 331 model. 


Turning now to the gauge boson content, one finds that the breaking process sketched above produces, in addition to the SM gauge bosons, an additional neutral $Z'$ boson
as well as singly charged $Y^{\pm}$ and doubly charged $Y^{\pm \pm}$ bosons. The photon is massless, as required, since a $U(1)$ remains unbroken while the
$W$ and $Z$ masses are at the weak scale, as required. Finally, all additional gauge bosons obtain their masses from the large VEV $v_{\sigma}$ 
and are therefore heavier. Assuming that this heavy VEV is much larger than the others, an interesting relation between the heavy gauge boson masses can be found, 
which allows to express the $Y^{\pm}$ mass in terms of the $Z'$ mass or vice versa:
\begin{equation}\label{massrel}
M_{Y^\pm}^2=\frac{3}{4} \frac{(1-4 \sin^2 \theta_W)}{(\cos^2 \theta_W)} M_{Z'}^2,
\end{equation}
while $M_{Y^{\pm}}=M_{Y^{\pm\pm}}$.
We will be more explicitely concerned with the respective fermion couplings below, but let us
already here state that in the quark sector the new charged gauge bosons always transmit a coupling from a SM light quark to an additional heavy one, and therefore do not
modify the amplitudes of low energy observables at tree level\footnote{This is no longer true in the lepton sector, where for example a new tree level diagram to muon decay 
modifies the Fermi coupling constant \cite{PSS}.}. The neutral $Z'$, on the other hand, also has couplings to light quarks only, and does give this kind of 
contributions.
In the decay $B \to X_s \gamma$, however, both charged and neutral gauge bosons are equally important, so that information on the charged bosons may be obtained.

After these more general remarks, let us now show explicitely the respective fermion couplings. Concerning the neutral currents, the Lagrangian for the above-mentioned FCNCs 
at tree level is given by   
\begin{equation}\label{FCNC}
\mathcal{L}_{FCNC}=\frac{g c_W}{\sqrt{3} \sqrt{1 - 4 s_W^2}}[\overline{u}\gamma_\mu\gamma_L
U_L^\dagger\pmatr{0&&\cr&0&\cr&&1}U_Lu+\overline{d}\gamma_\mu\gamma_L
\tilde V_L^\dagger\pmatr{0&&\cr&0&\cr&&1}\tilde V_Ld] {Z'}^{\mu}\,.
\end{equation}
{The explicit couplings to fermions and a possible parameterization for the new mixing
  matrices $U_L$ and $\tilde V_L$, that diagonalize the up and down-type Yukawa couplings, respectively, have been given in  \cite{PSS}.
These obey 
\begin{equation}
U_L^{\dagger} \tilde V_L = V_{\rm CKM},
\label{CKM} 
\end{equation}}
 and the notation of (\ref{FCNC}) and (\ref{CKM}) is to be understood as in \cite{PSS}, so that the tilde distinguishes between the SM CKM matrix $V_{\rm CKM}$ and the mixing matrix for the down-type quarks.

\begin{figure}
\begin{center}

\subfigure[]{\includegraphics[height=2.95cm]{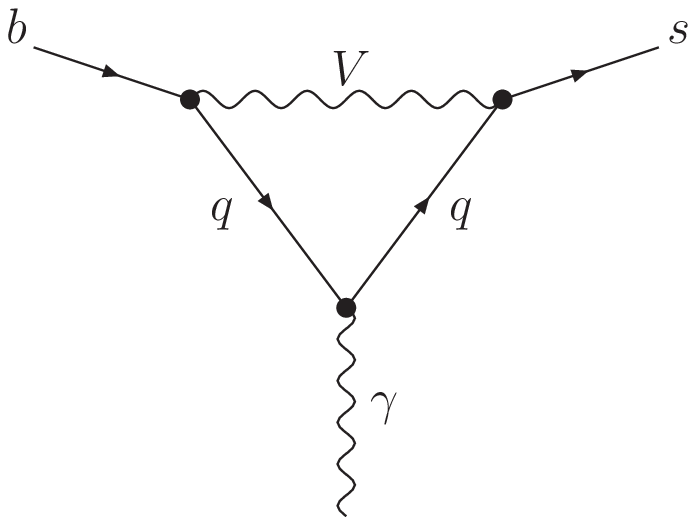}\label{Diag1}}
\subfigure[]{\includegraphics[height=2.95cm]{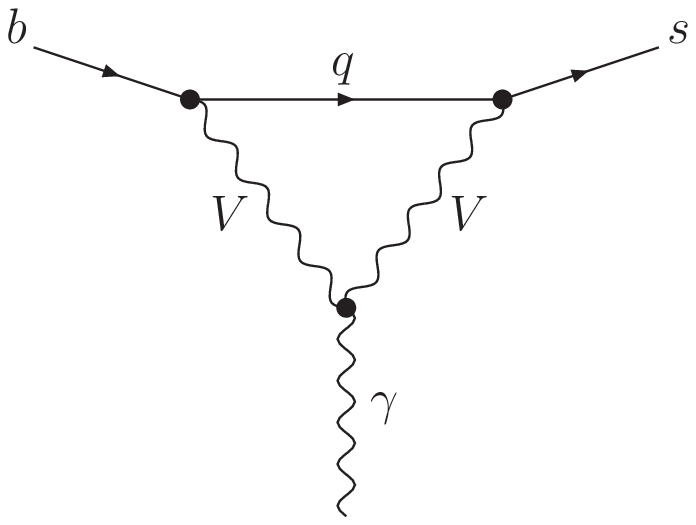}\label{Diag2}}
\subfigure[]{\includegraphics[height=2.95cm]{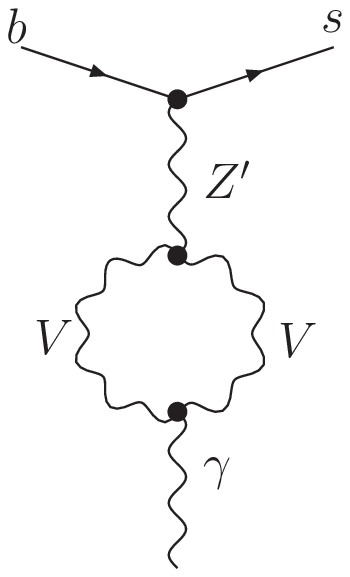}\label{Diag3}}
\subfigure[]{\includegraphics[height=2.95cm]{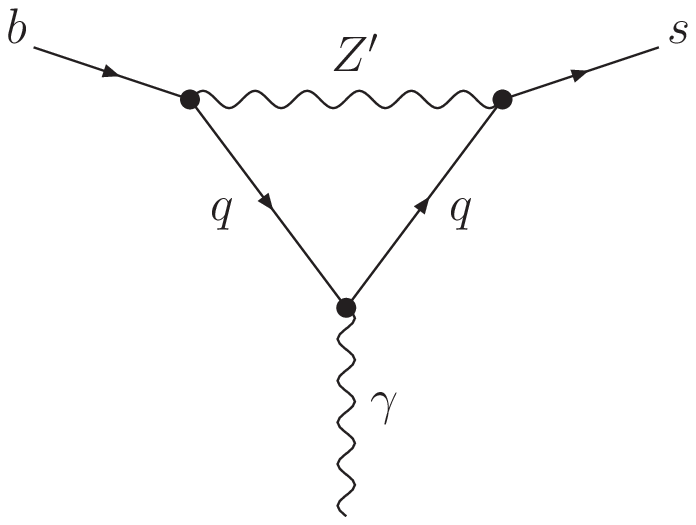}\label{Diag4}}
 \caption{{New magnetic penguin diagrams contributing to the decay $B \to X_s \gamma$. $V$ denotes the heavy charged gauge bosons $Y^\pm, Y^{\pm\pm}$.} We do not show {self-energies and} the diagrams involving Goldstone bosons.}
 \label{Fig1}
\end{center}
\end{figure}

Additionally, the charged current vertices in this basis are then
\begin{displaymath}\label{CCWvert}
J_{W^+}^\mu=\overline{u}\gamma^\mu\gamma_L U_L^{\dagger} \tilde V d  =\overline{u}\gamma^\mu\gamma_LV_{CKM}d
\end{displaymath}
\begin{eqnarray}\label{CCvert}
J_{Y^+}^\mu&=&\overline{d}\gamma^\mu\gamma_L \tilde V^\dagger
\pmatr{1&0\cr0&1\cr0&0}D+\overline{T}\gamma^\mu\gamma_L\pmatr{0&0&1}
U_Lu\nonumber\\
J_{Y^{++}}^\mu&=&\overline{u}\gamma^\mu\gamma_LU_L^\dagger
\pmatr{1&0\cr0&1\cr0&0}D-\overline{T}\gamma^\mu\gamma_L\pmatr{0&0&1}\tilde V d \, .
\label{eq:qcc}
\end{eqnarray}
From these equations, it becomes obvious that there can be new penguin diagrams in
the decay $B \to X_s \gamma$ containing  heavy quarks and heavy gauge bosons
in the loop. We will discuss the result for these diagrams, shown in Fig.~\ref{Fig1}, along with all other new contributions in the next section. Concerning the notation, 
here and 
in the following we denote by $V$ the heavy charged gauge bosons.
 
In the Higgs sector the symmetry breaking structure leads to a number of charged and neutral Higgs fields,
which in principle also transmit flavor changing interactions. Considering the heavy top quark 
coupling only, the quark sector can be identified with a modified 2 Higgs
doublet model (2HDM) type II, in which only the third generation diagrams
are taken into account\footnote{Note also that due to the exchange of the top and bottom quark in the third quark triplet, the couplings are accordingly exchanged.}.
This corresponds to \cite{Agrawal:1995vp}, where the Higgs sector
  has been approximated in this way. Note that these kind of terms do not appear 
in FCNC processes mediated by a tree level $Z'$, since there the top  Yukawa does not appear. However, for $B \to X_s \gamma$ they should be 
added. 

\section{Present Situation of the Decay $B \to X_s \gamma$}\label{sec:obs}
We begin with several general remarks concerning the decay $B \to X_s \gamma $, while the interested reader may consult \cite{Haisch:2007ic} for a very recent and 
more elaborate discussion of the theoretical and experimental background. 
In general, theoretical interest in this decay stems from the following 
features:
\begin{itemize}
\item Being an inclusive process, it can be calculated much more reliably than exclusive processes generally can be. Using the heavy quark expansion,
the calculation can be performed at the partonic level, while additional corrections are suppressed by terms of $\mathcal{O}(1/m_b)$.
\item The decay is by now very precisely measured. In addition to the increasingly precise SM calculation, it therefore offers a very good test of the SM
or a nice tool to search for physics beyond it.
\end{itemize}
Theoretically the decay $B \to X_s \gamma$ has now been calculated completely up to NLL 
precision, and recently a first estimate of the value at NNLL precision has been obtained \cite{Misiak:2006zs}, after a great effort of many groups to calculate the 
different contributions required \cite{BSGNNLO}. It reads:
\begin{equation}
Br(B \to X_s \gamma)|_{\rm NNLL}=(3.15 \pm 0.23) \times 10^{-4}
\end{equation}
This effort of the NNLL calculation had become necessary, due to a large uncertainty of the NLL calculation resulting from uncertainties in the charm mass 
renormalization scheme and scale.

On the other hand, the current experimental average is combined from the measurements of BaBar, Belle, CLEO and others, using inclusive and exclusive methods. Using 
a photon energy cut of $E_{\gamma} > 1.6$, HFAG gives an average of \cite{BBpage}
\begin{equation}
Br(B \to X_s \gamma)|_{\rm Exp}=(3.55 \pm 0.24^{+0.09}_{-0.10}\pm 0.03 ) \times 10^{-4} \,,
\end{equation} 
where the errors include also those from the extrapolation to lower energies.
Therefore, the experimental value is somewhat higher than the theoretical one, but still quite well compatible.
 
Let us also note here that the decay $B \to X_s \gamma$ has been analyzed in various models beyond the SM, among these the 2HDM type II \cite{2HDMpapers}, 
the general as well as the minimal flavor violating (MFV) minimal supersymmetric SM (MSSM) \cite{bsgSUSY},
 models with one \cite{Haisch:2007vb} or two \cite{Freitas:2008vh} universal extra dimensions (UED) and the littlest Higgs model with \cite{BBPTUW} and without 
\cite{Huo:2003vd} T parity. Some of these analyses have been performed at the NLO level, but most of them are done at LO order only. A comparison and compilation of the
main results can be found in
\cite{Haisch:2007ic}.

\subsection{New Contributions to $B \to X_s \gamma$ in the Minimal 331
  Model}

In this section, we list all the different contributions to the decay
$B \to X_s \gamma$ that modify the prediction of the 331 model with respect to the SM, and comment briefly
on the cancellation of divergences. {All expressions are explicitely given in
\cite{Agrawal:1995vp}.
Concentrating on the operators that are relevant for $B \to X_s \gamma$,
we have also confirmed the calculation of \cite{Agrawal:1995vp}.}

There are three different possibilities for the additional particles to show up in the decay $B \to X_s \gamma${. First of all, there are penguin diagrams involving the new charged gauge bosons, shown in the first two diagrams of Fig.~\ref{Fig1}.} For an arbitrary charged gauge boson and quark, the general structure, after calculating all penguin and the self energy diagrams required and summing over all of these, is a well known 
generalization of the Inami Lim Functions. In this calculation, a divergent term remains, which additionally violates gauge invariance. Within the SM, this term is 
canceled by the GIM mechanism when summing over all quark
flavors. {In the 331 
model, this singularity is not removed
entirely even then, due to the different charge assignment of the quarks. In this
case, the cancellation is achieved \cite{Agrawal:1994pf,Agrawal:1995vp} by adding the $Z' -\gamma$ mixing diagram shown in Fig.~\ref{Diag3}.}
Therefore, the very same mechanism that generated the FCNCs at tree level can potentially also enhance them at loop level, in particular, if the GIM mechanism is
as effective in the SM as it is in $B \to X_s \gamma$.

Next, there are also the $Z'$  penguin diagrams shown in Fig.~\ref{Diag4}.
To simplify the 
expressions, we assume that all quark masses vanish (in this case, there are down type quarks also in the loop). The unitarity of the mixing matrix $\tilde V$ then 
allows to sum up the terms from the $d$, $c$, and $b$ quarks into a simple form
without any leftover divergence. 

Finally, there are contributions from the Higgs sector of the model. Calculating these explicitely results in extremely complicated expressions,
due to the elaborate structure of the Higgs mass terms.
Resorting to the simplifications
 mentioned above, one can describe the Higgs sector by the 2HDM, where
 the corresponding expressions are well known \cite{2HDMpapers}.

\subsection{Initial Conditions and Renormalization Group Analysis}

The standard procedure of calculating any weak decay is the framework of a weak effective Hamiltonian. Here, the separation of scales is achieved by integrating out 
all heavy degrees of freedom, and casting the Hamiltonian into an effective form:
\begin{equation}
{\cal H}_{\rm eff}=-2\sqrt{2}G_F\sum_i C_i(\mu)O_i(\mu)\ .
\label{eq:heff331}
\end{equation}
The sum extends over all operators that can appear at a given scale. Within the SM, the masses of the 
heavy particles are all of the order of the weak scale, at which initial conditions are then calculated. Next, the QCD corrections arising from the renormalization group 
running further enhance the branching fraction of $B \to X_s \gamma$ by about a factor of 3, due to the effectiveness of the GIM suppression in this case. Within the
the 331 model, the QCD corrections are not expected to be as important \cite{Agrawal:1995vp}, but we take them into account for completeness. We use the leading order 
formulae for the RG running, with the anomalous dimensions as given in \cite{Agrawal:1995vp}, where they have been extended from the
SM to include the additional operators present in the 331 case. Turning to the matching conditions in the 331 model, 
there are evidently now additional scales in the problem. In principle, there is the scale of the $Z'$ boson mass, as well as that of the masses of the charged gauge bosons.
The QCD running between these two scales does not modify the values of the coefficients by much, and we therefore integrate out all heavy particles at the lower scale
$M_Y$. These are, in addition to the heavy gauge bosons, also the heavy quarks, while the Higgs sector is added at lower energies.

These initial conditions are run down to the weak scale, where several changes occur: First, the top quark is integrated out, and along with it the operators 
{in which the top quark appears. These are replaced by the standard operators $Q_{1/2}$ involving the charm quark.} Next, we should here add the SM matching conditions as well
as those from the 2HDM. 

Finally, from these coefficients, the branching ratio of $B \to X_s \gamma$ is calculated by
\begin{equation}
{\Gamma(b\to s\gamma)\over\Gamma(b\to ce\overline{\nu}_e)}
={|V_{ts}^*V_{tb}^{\vphantom{*}}|^2\over|V_{cb}|^2}{6\alpha\over\pi C}
\left(|C_7(\mu_b)|^2+N(E_0)\right)\ ,
\label{eq:rate}
\end{equation}
in the notation of \cite{Misiak:2006ab}, where $C=|V_{ub}/V_{cb}|^2 \Gamma(b\to ce\overline{\nu}_e)/\Gamma(b\to ue\overline{\nu}_e)=0.58 \pm 0.016$
and $N(E_0)=0.0031$ are the nonperturbative corrections (we take $Br(b\to ce\overline{\nu}_e)=0.106$).
Using the SM expression for $C_7$ given above leads to the LO value of the branching fraction, which, numerically, does not agree with the most recent theoretical value 
given in \cite{Misiak:2006zs}. To accommodate for the corresponding shift, we directly set the
SM part of $C_7$ to the NNLO value and go through the entire RGE procedure only for the new physics contributions.

\section{Numerical Analysis}\label{sec:numerics}
\subsection{Preliminaries}
We will now analyze numerically the new contributions to the decay $B
\to X_s \gamma$ and investigate whether additional bounds on the minimal 331 model can be obtained
from this mode and which of the parameters appearing have the strongest influence. 

As constraints we will use the existing data from $K$ and $B$ meson mixing, such as
 $\Delta M_{d/s}$, $\sin 2 \beta$, $\Delta M_K$ and $\eps_K$  whose theoretical expressions in the 331 model are given in \cite{PSS}. As numerical input we will take the tree level values of $|V_{us}|$, $|V_{ub}|$,
$|V_{cb}|$ given in \cite{PSS}, and $\gamma=(82\pm20)^\circ$.
Further, we follow the analysis of \cite{BBPTUW} and set all non-perturbative 
parameters to their central values and allow $\Delta M_K$, 
$\varepsilon_K$, $\Delta M_d$, and $\Delta M_s$ to differ from 
their experimental values by $\pm 50\%$, $\pm 40\%$, $\pm 40\%$ and $\pm 40\%$ 
 respectively. In the case of $\Delta M_s/\Delta M_d$ we
will choose $\pm 20\%$ as the error on the relevant parameter $\xi$ is
smaller than in the case of $\Delta M_d$ and $\Delta M_s$ separately. 
Similarly, we will absorb the SM uncertainties of $B \to X_s \gamma$, stemming mainly from the remaining scale uncertainty in the charm mass as well as the CKM factors,
into the experimental uncertainty by increasing it to $ \pm 15\%$ instead of the $ \pm8\%$ given above.

We then perform a scan over the parameters of the 331 model considering the following ranges:
\be\nonumber
m_T,m_S=250-1000\;{\rm GeV}, \qquad M_{H^+}=300-2000\;{\rm GeV}, \qquad
M_{Z^\prime}=1000-5000\;{\rm GeV}.
\ee
The three angles and the three phases of the new mixing matrix $\tilde V$ are
kept arbitrary. Further, the expressions for the 2HDM depend on $\tan \beta$, which we mainly restrict to values of $\tan \beta >1$ for reasons that will 
become evident during our analysis.
We observe that, in a parameterization of the $\tilde V$ matrix that keeps the real and imaginary 
parts of the relevant combinations of $V_{ij}^* V_{lm}$ as
independent, the only bound we need to consider will be the one coming from
$\Delta M_s$.

\subsection{Constraints from $B\to X_s \gamma$ Compared to Other Constraints}

\begin{figure}
\centerline{
{
\epsffile{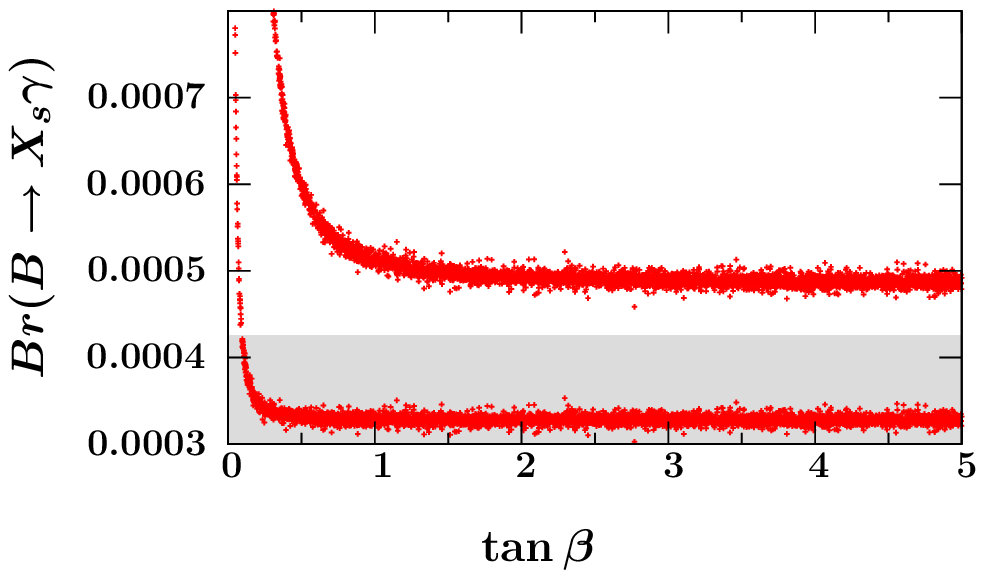}}}
\caption{The dependence of $Br(B \to X_s \gamma)$ on $\tan
\beta$. For values of $\tan \beta$ below 0.5 significant enhancements are
possible. The grey band indicates $Br(B \to X_s \gamma)|_{\rm Exp}$, with errors inflated as described in the text. The upper band shows the values for $M_{H^+}=300 \; \gev$, while the lower one shows 
those for $M_{H^+}=5000 \; \gev$}
\label{fig:tanbeta}%
\end{figure}

Let us first elaborate on the influence of the chosen values for $\tan \beta$.
Looking at Fig.~\ref{fig:tanbeta}, we observe that the value obtained for $B \to X_s \gamma$ is practically
independent of its numerical value as long as $\tan \beta > 2$. 
On the other hand, large values of the branching fraction can be obtained for smaller values of $\tan \beta$. 
This effect of the 2HDM is well known \cite{2HDMpapers}, and has most recently been investigated and 
numerically updated in \cite{Misiak:2006zs}. We refer the reader to the detailed discussion of $B \to X_s \gamma$ within the 2HDM given there,
and in the following fix $\tan \beta=2$, in order to show more clearly the additional effects of the 331 model.   
In this context, we would like to point out that, in the pure 2HDM, very small values of $\tan \beta$ are excluded by
other observables, such as electroweak precision tests. While the 331 model does no longer resemble the 2HDM when gauge bosons are 
included, we use only the larger values for $\tan \beta$.

\begin{figure}
\centerline{
{
\epsffile{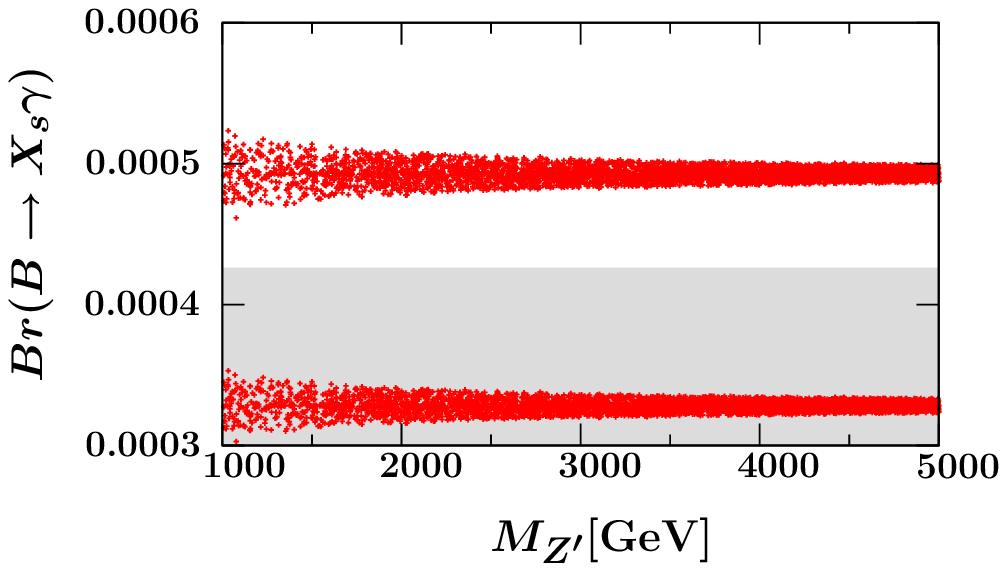}}}
\caption{$Br(B\to X_s \gamma)$ versus $M_{Z^\prime}$. The grey
band shows the experimental range with a 15 $ \%$ error. The upper
band corresponds to a Higgs mass of $M_{H^+}=300\gev$ and the lower one to  $M_{H^+}=2000\gev$.}
\label{fig:mZ}
\end{figure}

Next, we show in  Fig.~\ref{fig:mZ} the dependence of the branching fraction on the $Z'$ mass, where we 
separate the dependence on the charged 
Higgs mass by showing only the values obtained for the upper and lower bound on $M_{H^+}$, respectively. In any case, the upper line 
corresponds to the lower value of $M_{H^+}$, meaning that the branching fraction increases with decreasing Higgs mass.
The width of the
remaining bands corresponds then to the allowed range of $V_{32}^* V_{33}$ and the heavy quark masses.
Clearly, a variation in the charged Higgs mass has a much stronger influence on the value of the branching fraction, which
leads us to conclude that, as a whole, the 2HDM contributions vastly dominate over those from the gauge bosons, and that,
therefore, the signature of the minimal 331 model with respect to $B \to X_s \gamma$ is basically identical. On the other
hand, the bounds on the Higgs sector from the type II 2HDM can immediately be applied to the 331 model. Also, we can conclude
that the decay $B \to X_s \gamma$ would be well suited to explore the Higgs sector of the 331 model without additional 
pollution from gauge bosons, if the minimal 331 model should be established through other channels. These results represent the main conclusions of our analysis.

\begin{figure}
\centerline{
{
\epsffile{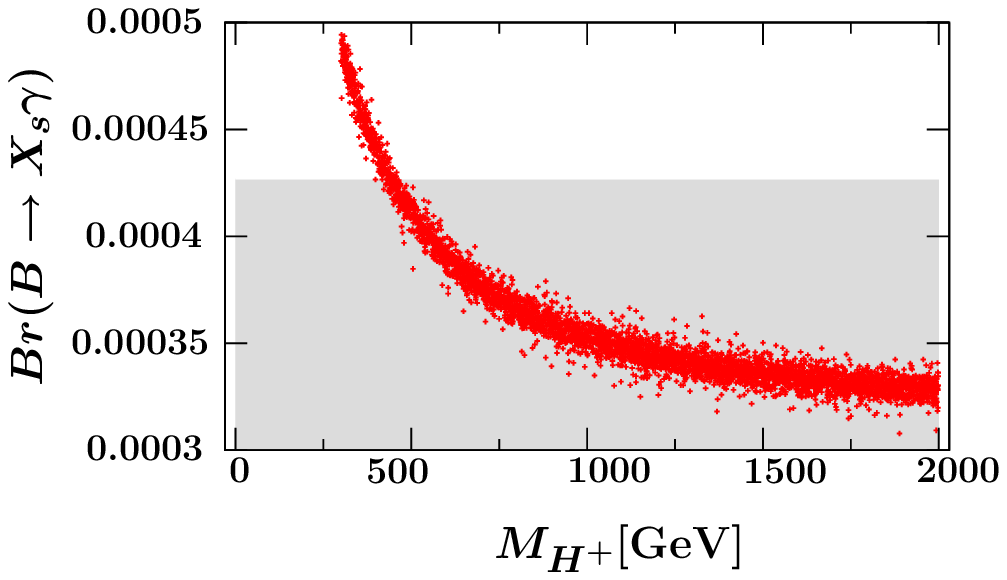}}
}
\caption{$Br(B\to X_s \gamma)$ versus $M_{H^+}$. The grey
band shows the experimental range with a 15 $ \%$ error. From the plot one finds a lower bound on
the charged Higgs mass of roughly 400 GeV.}
\label{fig:mH}
\end{figure}

In this context, we also show the corresponding dependence of the
branching fraction on the charged Higgs mass $M_{H^+}$ in Fig.~\ref{fig:mH}, which in principle allows us to read off a lower bound for $M_{H^+} \approx 400 ~\gev$. 
One should now take into account the NLO corrections within the SM and the 2HDM, which are known to be significant, in order to improve quantitatively on this bound.
However, since $B\to X_s \gamma$ has been studied elaborately, both in the SM and the 2HDM, in \cite{Misiak:2006zs}, including every known contributions
to this decay, we consider it beyond the scope of our paper to repeat this analysis and simply quote the lower bound of the Higgs mass as
\begin{equation}
M_{H^+} \geq 295 \gev \qquad (95 \%\ C.L.)\,.
\end{equation}

Interestingly, while the 331 model can, in principle, either enhance or suppress the branching fraction of 
$B \to X_s \gamma$, depending on the sign of the mixing matrix elements, the model as a whole predicts an enhancement of the branching fraction, due
to this strong dominance of the 2HDM model. This is, of course, rather fortunate considering the present experimental result.

{We will close this subsection and our numerical analysis with some very brief remarks concerning the possible influence on the rare decays and the other observables discussed in
\cite{PSS}.
However, after the findings of the last paragraph, one expects this influence to be completely negligible.
In any case, the most interesting observable
is the $B_s - \bar B_s$ mixing phase, which can be large. This statement is not altered by our analysis of
$B \to X_s \gamma$. 
More interestingly, a first measurement of this quantity has recently appeared \cite{Abazov:2007zj}, which now offers some
hint as to its size. As of now, this measurement is not precise enough to warrant a more detailed discussion, but it will be extremely interesting to 
follow the progress of the corresponding experiments.}

\begin{figure}
\centerline{
{
\epsffile{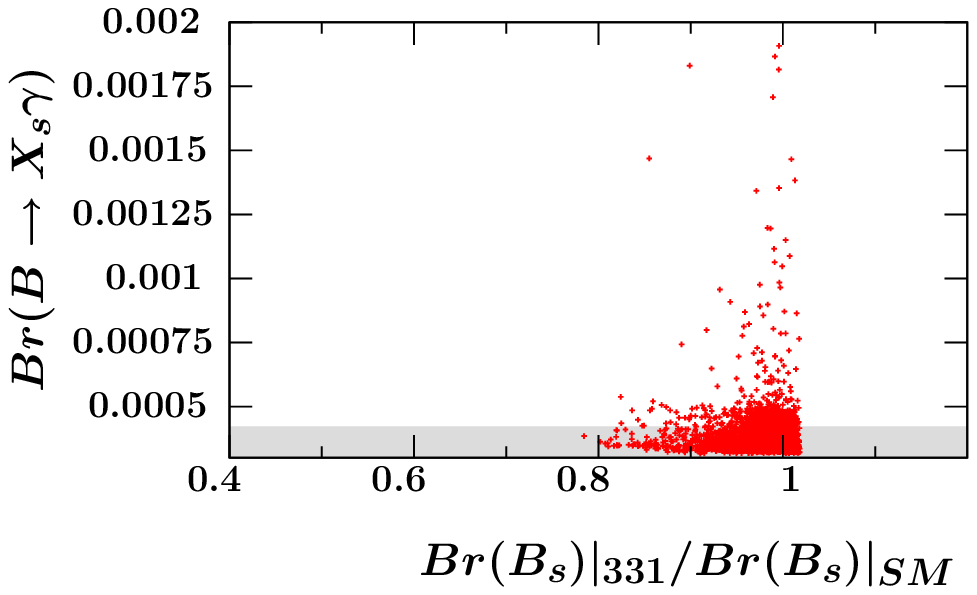}}}
\caption{Correlation of $B \to X_s \gamma$ with the deviation of $Br(B_s) \equiv Br(B_s \to \mu^+ \mu^-)$ from the SM. The experimental range of the former (given by grey band) does not further constrain the
range of the latter.}
\label{fig:Bmumu}
\end{figure}

All other statements of \cite{PSS} remain unaffected, in particular, one can still obtain sizeable effects in the
rare $K$ decays, while large effects in $B_{d/s} \to \mu^+ \mu^-$ seem excluded. To show this graphically, we show in Fig.~\ref{fig:Bmumu} the correlation between
the branching fraction of $B \to X_s \gamma$ with the ratio $Br(B_s \to \mu^+ \mu^-)|_{331}/Br(B_s \to \mu^+ \mu^-)|_{SM}$, where we observe that the grey band, 
representing the experimental range of $Br(B \to X_s \gamma)$, still allows for the entire range of $Br(B_s \to \mu^+ \mu^-)$.

\section{Conclusions}\label{sec:conclusions}
Using new data for both the decay $B \to X_s \gamma$ as well as for the other existing constraints available, and in view of recent theoretical progress, including a 
first NNLO estimate, we have reinvestigated the implications of the minimal 
331 model for the decay $B \to X_s \gamma$. In contrast to several other FCNC observables, that are affected at tree level by $Z'$ contributions within this model,
we are dealing here with a purely loop induced process, which, due to the breaking of the GIM mechanism, may still receive significant contributions.
We have in general retained the general feature of the more than ten year old analysis performed in \cite{Agrawal:1995vp},
which particularly concerns the performed simplifications within the Higgs sector. It is described here in terms of an effective 2HDM, so that the new contributions
to the branching fraction can be discussed in terms of two different parts: one originating from the extended Higgs sector and the other one from the additional gauge bosons and quarks. The latter are governed by the new mixing matrix $\tilde V$, that appears as a set of new parameters in the model, and is constrained by the 
existing bounds from FCNC processes. On the other hand, the 2HDM contribution is only constrained by the usual parameter constraints on the Higgs mass. 
The main new result of our analysis is the finding that the gauge contributions are now constrained so strongly that effectively one is left with the 2HDM ones. Thus, it is possible to obtain a lower bound on the charged Higgs mass, as obtained from a recent, more sophisticated 
analysis of the 2HDM. 
We therefore conclude that the decay $B \to X_s \gamma$ is a very useful tool to probe the Higgs sector
 of the minimal 331 model and is therefore complementary to other FCNC observables. 
This shows once again the power of the $B$ and $K$ meson systems in
obtaining information about models beyond the SM, when they are used together and combined.

\vspace{0.8cm}

\noindent
{\bf Acknowledgments}\\
\noindent
We would like to thank A.J. Buras for interesting discussions and U. Haisch for extremely valuable comments on the manuscript. 
F.S. acknowledges financial support from the Deutsche Forschungsgemeinschaft (DFG).
 Also, this work is supported in part by the
Cluster of Excellence "Origin and Structure of the Universe" and by the German Bundesministerium f\"ur Bildung und Forschung (BMBF) under contract 05HT6WOA.
S.U. would like to thank Gino Isidori and the LNF Spring Institute
2007 
for hospitality during the completion of this work.

\begin{appendix}
 \comment{
\section{Feynman Rules for Vertices}\label{sec:FR}
In this section, we list all the Feynman Rules relevant to our calculation. We define $P_L \equiv \frac{\gamma^{\mu}}{2} (1-\gamma^5)$ and 
$P_R \equiv \frac{\gamma^{\mu}}{2} (1+\gamma^5)$.
\indent

\subsubsection*{Quark - $Z'$ - vertices}
\indent

\begin{tabular}{ll}
\begin{picture}(150,70)(0,-10)
\Photon(10,0)(60,0){3}{4}
\Text(0,0)[c]{$Z'_{\mu}$}
\ArrowLine(60,0)(110,0)
\Text(120,0)[c]{$u_j$}
\ArrowLine(60,50)(60,0)
\Text(65,45)[l]{$u_i$}
\Vertex(60,0){2}
\end{picture}
&
\raisebox{35\unitlength}{
\begin{minipage}{5cm}
\lefteqn{
i \: \frac{g c_W}{\sqrt{3} \sqrt{1 - 4 s_W^2}} \: U_{3 U_i} U_{3 U_j}^\ast  \: P_L
\qquad i,j=1,2,3 \,, i \neq j}
\end{minipage}
}
\end{tabular}

\begin{tabular}{ll}
\begin{picture}(150,70)(0,-10)
\Photon(10,0)(60,0){3}{4}
\Text(0,0)[c]{$Z'_{\mu}$}
\ArrowLine(60,0)(110,0)
\Text(120,0)[c]{$u_i$}
\ArrowLine(60,50)(60,0)
\Text(65,45)[l]{$u_i$}
\Vertex(60,0){2}
\end{picture}
&
\raisebox{35\unitlength}{
\begin{minipage}{5cm}
\lefteqn{
 \frac{-i g }{\sqrt{3} c_W \sqrt{1 - 4 s_W^2}} \: \left( (1/2- s_W^2-c_W^2 U_{3 i} U_{3 i}^\ast) P_L - \:2 s_W^2 P_R \right) }
\end{minipage}
}
\end{tabular}


\begin{tabular}{ll}
\begin{picture}(150,70)(0,-10)
\Photon(10,0)(60,0){3}{4}
\Text(0,0)[c]{$Z'_{\mu}$}
\ArrowLine(60,0)(110,0)
\Text(120,0)[c]{$d_j$}
\ArrowLine(60,50)(60,0)
\Text(65,45)[l]{$d_i$}
\Vertex(60,0){2}
\end{picture}
&
\raisebox{35\unitlength}{
\begin{minipage}{5cm}
\lefteqn{
i \: \frac{g c_W}{\sqrt{3} \sqrt{1 - 4 s_W^2}} \: \tilde V_{3 d_i} \tilde V_{3 d_j}^\ast \: P_L
\qquad i,j=1,2,3 \,, i \neq j}
\end{minipage}
}
\end{tabular}

\begin{tabular}{ll}
\begin{picture}(150,70)(0,-10)
\Photon(10,0)(60,0){3}{4}
\Text(0,0)[c]{$Z'_{\mu}$}
\ArrowLine(60,0)(110,0)
\Text(120,0)[c]{$d_i$}
\ArrowLine(60,50)(60,0)
\Text(65,45)[l]{$d_i$}
\Vertex(60,0){2}
\end{picture}
&
\raisebox{35\unitlength}{
\begin{minipage}{5cm}
\lefteqn{
\frac{-i g}{\sqrt{3} c_W \sqrt{1 - 4 s_W^2}} \left( (1/2- s_W^2 - c_W^2 \tilde V_{3i} \tilde V_{3i}^*) \: P_L + s_W^2\:P_R \right)}
\end{minipage}
}
\end{tabular}


\begin{tabular}{ll}
\begin{picture}(150,70)(0,-10)
\Photon(10,0)(60,0){3}{4}
\Text(0,0)[c]{$Z'_{\mu}$}
\ArrowLine(60,0)(110,0)
\Text(120,0)[c]{$T$}
\ArrowLine(60,50)(60,0)
\Text(65,45)[l]{$T$}
\Vertex(60,0){2}
\end{picture}
&
\raisebox{35\unitlength}{
\begin{minipage}{5cm}
\lefteqn{
-i \: \frac{g}{\sqrt{3} c_W \sqrt{1 - 4 s_W^2}} \:  \left( (1-6 s_W^2) P_L - 5 s_W^2 P_R \right)
}
\end{minipage}
}
\end{tabular}

\begin{tabular}{ll}
\begin{picture}(150,70)(0,-10)
\Photon(10,0)(60,0){3}{4}
\Text(0,0)[c]{$Z'_{\mu}$}
\ArrowLine(60,0)(110,0)
\Text(120,0)[c]{$D_i$}
\ArrowLine(60,50)(60,0)
\Text(65,45)[l]{$D_i$}
\Vertex(60,0){2}
\end{picture}
&
\raisebox{35\unitlength}{
\begin{minipage}{5cm}
\lefteqn{
i \: \frac{g}{\sqrt{3} c_W \sqrt{1 - 4 s_W^2}}  \: \left( (1-5 s_W^2) P_L + 4 s_W^2 P_R \right) \qquad i = 1,2
}
\end{minipage}
}
\end{tabular}

\subsubsection*{Lepton - $Y^\pm$ - vertices}
\indent

\begin{tabular}{ll}
\begin{picture}(150,70)(0,-10)
\Photon(10,0)(60,0){3}{4}
\Text(0,0)[c]{$Y_{\mu}$}
\ArrowLine(60,0)(110,0)
\Text(120,0)[c]{$l^C$}
\ArrowLine(60,50)(60,0)
\Text(65,45)[l]{$\nu_l$}
\Vertex(60,0){2}
\end{picture}
&
\raisebox{35\unitlength}{
\begin{minipage}{5cm}
\lefteqn{
i \: \frac{g}{\sqrt{2}} \: P_L
}
\end{minipage}
}
\end{tabular}

\subsubsection*{Lepton - $Y^{\pm\pm}$ - vertices}
\indent

\begin{tabular}{ll}
\begin{picture}(150,70)(0,-10)
\Photon(10,0)(60,0){3}{4}
\Text(0,0)[c]{$Y_{\mu}$}
\ArrowLine(60,0)(110,0)
\Text(120,0)[c]{$l^C$}
\ArrowLine(60,50)(60,0)
\Text(65,45)[l]{$l$}
\Vertex(60,0){2}
\end{picture}
&
\raisebox{35\unitlength}{
\begin{minipage}{5cm}
\lefteqn{
-i \: \frac{g}{\sqrt{2}} \: P_L
}
\end{minipage}
}
\end{tabular}
}
\end{appendix}


\begin{thebibliography}{999}

\bibitem{Frampton:1992wt}
  P.~H.~Frampton,
  Phys.\ Rev.\ Lett.\  {\bf 69} (1992) 2889.

\bibitem{Pisano:1991ee}
  F.~Pisano and V.~Pleitez,
  Phys.\ Rev.\ D {\bf 46} (1992) 410;
  R.~Foot, O.~F.~Hernandez, F.~Pisano and V.~Pleitez,
  Phys.\ Rev.\ D {\bf 47} (1993) 4158.

\bibitem{Liu:1993gy}
  J.~T.~Liu and D.~Ng,
  Phys.\ Rev.\ D {\bf 50} (1994) 548.


\bibitem{GomezDumm:1994tz}
  D.~Gomez Dumm, F.~Pisano and V.~Pleitez,
  Mod.\ Phys.\ Lett.\ A {\bf 9} (1994) 1609.

\bibitem{Rodriguez:2004mw}
  J.~A.~Rodriguez and M.~Sher,
  Phys.\ Rev.\ D {\bf 70} (2004) 117702.


\bibitem{PSS}
  C.~Promberger, S.~Schatt and F.~Schwab,
  Phys.\ Rev.\  D {\bf 75} (2007) 115007.

\bibitem{Martinez:2008jj}
  R.~Martinez and F.~Ochoa,
  arXiv:0802.0309 [hep-ph].


\bibitem{Agrawal:1995vp}
  J.~Agrawal, P.~H.~Frampton and J.~T.~Liu,
  Int.\ J.\ Mod.\ Phys.\ A {\bf 11} (1996) 2263
%

\bibitem{Agrawal:1994pf}
  J.~Agrawal and P.~H.~Frampton,
  Nucl.\ Phys.\  B {\bf 419} (1994) 254.

\bibitem{Koppenburg:2004fz}
  P.~Koppenburg {\it et al.}  [Belle Collaboration],
  Phys.\ Rev.\ Lett.\  {\bf 93} (2004) 061803.

\bibitem{Aubert:2005cua}
B.~Aubert {\it et al.}  [BaBar Collaboration],
  Phys.\ Rev.\  D {\bf 72} (2005) 052004.


\bibitem{Aubert:2006gg}
  B.~Aubert {\it et al.}  [BaBar Collaboration],
  Phys.\ Rev.\ Lett.\  {\bf 97} (2006) 171803.

\bibitem{Liu:1994rx}
  J.~T.~Liu,
  Phys.\ Rev.\ D {\bf 50} (1994) 542




\bibitem{Ng:1992st}
  D.~Ng,
  Phys.\ Rev.\ D {\bf 49} (1994) 4805.

\bibitem{Foot:1994ym}
  R.~Foot, H.~N.~Long and T.~A.~Tran,
  Phys.\ Rev.\ D {\bf 50} (1994) 34;
  H.~N.~Long,
  Phys.\ Rev.\ D {\bf 53} (1996) 437;
  H.~N.~Long,
  Phys.\ Rev.\ D {\bf 54} (1996) 4691;


\bibitem{Montero:1992jk}
  J.~C.~Montero, F.~Pisano and V.~Pleitez,
  Phys.\ Rev.\ D {\bf 47} (1993) 2918.
\bibitem{Long:1999ij}
  H.~N.~Long and V.~T.~Van,
  J.\ Phys.\ G {\bf 25} (1999) 2319;


\bibitem{Gutierrez:2004sb}
  D.~A.~Gutierrez, W.~A.~Ponce and L.~A.~Sanchez,
  Eur.\ Phys.\ J.\ C {\bf 46} (2006) 497.




\bibitem{Diaz:2004fs}
  R.~A.~Diaz, R.~Martinez and F.~Ochoa,
  Phys.\ Rev.\ D {\bf 72} (2005) 035018;
  F.~Ochoa and R.~Martinez,
  Phys.\ Rev.\ D {\bf 72} (2005) 035010;
  F.~Ochoa and R.~Martinez,
  hep-ph/0508082;
  A.~Carcamo, R.~Martinez and F.~Ochoa,
  Phys.\ Rev.\ D {\bf 73} (2006) 035007.


\bibitem{Okamoto:1999cf}
  Y.~Okamoto and M.~Yasue,
  Phys.\ Lett.\ B {\bf 466} (1999) 267

\bibitem{Kitabayashi:2000nq}
  T.~Kitabayashi and M.~Yasue,
  Phys.\ Rev.\ D {\bf 63} (2001) 095002.

\bibitem{Tully:2000kk}
  M.~B.~Tully and G.~C.~Joshi,
  Phys.\ Rev.\ D {\bf 64} (2001) 011301



\bibitem{Montero:2001ts}
  J.~C.~Montero, C.~A.~De S. Pires and V.~Pleitez,
  Phys.\ Rev.\ D {\bf 65} (2002) 095001

\bibitem{Cortez:2005cp}
  N.~V.~Cortez and M.~D.~Tonasse,
  Phys.\ Rev.\ D {\bf 72} (2005) 073005.

\bibitem{Duong:1993zn}
  T.~V.~Duong and E.~Ma,
  Phys.\ Lett.\ B {\bf 316} (1993) 307.

\bibitem{Montero:2000ng}
  J.~C.~Montero, V.~Pleitez and M.~C.~Rodriguez,
  Phys.\ Rev.\ D {\bf 65} (2002) 035006.

\bibitem{Montero:2004uy}
  J.~C.~Montero, V.~Pleitez and M.~C.~Rodriguez,
  Phys.\ Rev.\ D {\bf 70} (2004) 075004.

\bibitem{Ponce:2002sg}
  W.~A.~Ponce, Y.~Giraldo and L.~A.~Sanchez,
  Phys.\ Rev.\  D {\bf 67} (2003) 075001;
  P.~V.~Dong, H.~N.~Long, D.~T.~Nhung and D.~V.~Soa,
  Phys.\ Rev.\  D {\bf 73} (2006) 035004.


\bibitem{Anh:2000bs}
  N.~T.~Anh, N.~A.~Ky and H.~N.~Long,
  Int.\ J.\ Mod.\ Phys.\ A {\bf 16} (2001) 541.


\bibitem{Diaz:2003dk}
  R.~A.~Diaz, R.~Martinez and F.~Ochoa,
  Phys.\ Rev.\ D {\bf 69} (2004) 095009.





\bibitem{LP}
 A.~G.~Dias, R.~Martinez and V.~Pleitez,
  Eur.\ Phys.\ J.\ C {\bf 39} (2005) 101.


\bibitem{Haisch:2007ic}
  U.~Haisch,
  arXiv:0706.2056 [hep-ph].

\bibitem{Misiak:2006zs}
  M.~Misiak {\it et al.},
  Phys.\ Rev.\ Lett.\  {\bf 98} (2007) 022002.

\bibitem{BSGNNLO}
  K.~Bieri, C.~Greub and M.~Steinhauser,
  Phys.\ Rev.\  D {\bf 67}, 114019 (2003);
  M.~Misiak and M.~Steinhauser,
  Nucl.\ Phys.\  B {\bf 683}, 277 (2004);
  M.~Gorbahn and U.~Haisch,
  Nucl.\ Phys.\  B {\bf 713}, 291 (2005);
  M.~Gorbahn, U.~Haisch and M.~Misiak,
  Phys.\ Rev.\ Lett.\  {\bf 95}, 102004 (2005);
  K.~Melnikov and A.~Mitov,
  Phys.\ Lett.\  B {\bf 620}, 69 (2005);
  I.~Blokland, A.~Czarnecki, M.~Misiak, M.~Slusarczyk and F.~Tkachov,
  Phys.\ Rev.\  D {\bf 72}, 033014 (2005);
  H.~M.~Asatrian, A.~Hovhannisyan, V.~Poghosyan, T.~Ewerth, C.~Greub and T.~Hurth,
  Nucl.\ Phys.\  B {\bf 749}, 325 (2006);
  H.~M.~Asatrian, T.~Ewerth, A.~Ferroglia, P.~Gambino and C.~Greub,
  Nucl.\ Phys.\  B {\bf 762}, 212 (2007);
  M.~Czakon, U.~Haisch and M.~Misiak,
  JHEP {\bf 0703}, 008 (2007);
  M.~Misiak and M.~Steinhauser,
  Nucl.\ Phys.\  B {\bf 764}, 62 (2007).

\bibitem{2HDMpapers}
  M.~Ciuchini, G.~Degrassi, P.~Gambino and G.~F.~Giudice,
  Nucl.\ Phys.\  B {\bf 527} (1998) 21;
  F.~Borzumati and C.~Greub,
  Phys.\ Rev.\  D {\bf 58} (1998) 074004;
  

\bibitem{bsgSUSY}
  M.~Ciuchini, G.~Degrassi, P.~Gambino and G.~F.~Giudice,
  Nucl.\ Phys.\  B {\bf 534}, 3 (1998);
  F.~Borzumati, C.~Greub and Y.~Yamada,
  Phys.\ Rev.\  D {\bf 69}, 055005 (2004);
  G.~Degrassi, P.~Gambino and P.~Slavich,
  Phys.\ Lett.\  B {\bf 635}, 335 (2006);
  G.~Degrassi, P.~Gambino and G.~F.~Giudice,
  JHEP {\bf 0012}, 009 (2000);
  M.~S.~Carena, D.~Garcia, U.~Nierste and C.~E.~M.~Wagner,
  Phys.\ Lett.\  B {\bf 499}, 141 (2001);
  A.~J.~Buras, P.~H.~Chankowski, J.~Rosiek and L.~Slawianowska,
  Nucl.\ Phys.\  B {\bf 659}, 3 (2003);
  A.~Freitas, E.~Gasser and U.~Haisch,
  Phys.\ Rev.\  D {\bf 76}, 014016 (2007);
  G.~D'Ambrosio, G.~F.~Giudice, G.~Isidori and A.~Strumia,
  Nucl.\ Phys.\  B {\bf 645}, 155 (2002);
  F.~Borzumati, C.~Greub, T.~Hurth and D.~Wyler,
  Phys.\ Rev.\  D {\bf 62}, 075005 (2000);
  T.~Besmer, C.~Greub and T.~Hurth,
  Nucl.\ Phys.\  B {\bf 609}, 359 (2001);
  M.~Ciuchini, E.~Franco, A.~Masiero and L.~Silvestrini,
  Phys.\ Rev.\  D {\bf 67}, 075016 (2003)
  [Erratum-ibid.\  D {\bf 68}, 079901 (2003)];
  J.~Foster, K.~i.~Okumura and L.~Roszkowski,
  JHEP {\bf 0508}, 094 (2005);
  J.~Foster, K.~i.~Okumura and L.~Roszkowski,
  Phys.\ Lett.\  B {\bf 641}, 452 (2006);
  M.~Ciuchini, A.~Masiero, P.~Paradisi, L.~Silvestrini, S.~K.~Vempati and O.~Vives,
  arXiv:hep-ph/0702144.



\bibitem{Haisch:2007vb}
  K.~Agashe, N.~G.~Deshpande and G.~H.~Wu,
  Phys.\ Lett.\  B {\bf 514}, 309 (2001);
  A.~J.~Buras, A.~Poschenrieder, M.~Spranger and A.~Weiler,
  Nucl.\ Phys.\  B {\bf 678}, 455 (2004);
  U.~Haisch and A.~Weiler,
  Phys.\ Rev.\  D {\bf 76} (2007) 034014.

\bibitem{Freitas:2008vh}
  A.~Freitas and U.~Haisch,
  arXiv:0801.4346 [hep-ph].


\bibitem{BBPTUW}
  M.~Blanke, A.~J.~Buras, A.~Poschenrieder, C.~Tarantino, S.~Uhlig and A.~Weiler,
  JHEP {\bf 0612} (2006) 003;
  M.~Blanke, A.~J.~Buras, S.~Recksiegel, C.~Tarantino and S.~Uhlig,
  JHEP {\bf 0706} (2007) 082. 

\bibitem{Huo:2003vd}
  W.~j.~Huo and S.~h.~Zhu,
  Phys.\ Rev.\  D {\bf 68} (2003) 097301;
  A.~J.~Buras, A.~Poschenrieder, S.~Uhlig and W.~A.~Bardeen,
  JHEP {\bf 0611}, 062 (2006).

\bibitem{Misiak:2006ab}
  M.~Misiak and M.~Steinhauser,
  Nucl.\ Phys.\  B {\bf 764} (2007) 62.


\bibitem{BBpage}
The Heavy Flavor Averaging Group (HFAG),\\
http://www.slac.stanford.edu/xorg/hfag/.

\bibitem{CKM2005}
  E.~Blucher {\it et al.},
  arXiv:hep-ph/0512039.


\bibitem{PDG}
  S.~Eidelman {\it et al.}  [Particle Data Group],
  Phys.\ Lett.\ B {\bf 592} (2004) 1.
\bibitem{Hashimoto}
  S.~Hashimoto,
  Int.\ J.\ Mod.\ Phys.\ A {\bf 20} (2005) 5133.


\bibitem{UTFIT}
  M.~Bona {\it et al.}  [UTfit Collaboration],
  arXiv:hep-ph/0509219; arXiv:hep-ph/0605213; 
http://utfit.roma1.infn.it.


\bibitem{CDFnew}
A.~Abulencia {\it et al.}  [CDF Collaboration],
  Phys.\ Rev.\ Lett.\  {\bf 97} (2006) 242003.

\bibitem{D0}
  V.~M.~Abazov {\it et al.}  [D0 Collaboration],
  Phys.\ Rev.\ Lett.\  {\bf 97} (2006) 021802
  [arXiv:hep-ex/0603029].


\bibitem{eta1}
  S.~Herrlich and U.~Nierste,
  Nucl.\ Phys.\ B {\bf 419} (1994) 292.

\bibitem{eta3}
  S.~Herrlich and U.~Nierste,
  Phys.\ Rev.\ D {\bf 52} (1995) 6505;
  Nucl.\ Phys.\ B {\bf 476} (1996) 27.

\bibitem{eta2B}
  A.~J.~Buras, M.~Jamin and P.~H.~Weisz,
  Nucl.\ Phys.\ B {\bf 347} (1990) 491.
  J.~Urban, F.~Krauss, U.~Jentschura and G.~Soff,
  Nucl.\ Phys.\ B {\bf 523} (1998) 40.




\bibitem{Abazov:2007zj}
  V.~M.~Abazov {\it et al.}  [D0 Collaboration],
  hep-ex/0702030.



































.







































%
%
%
\end{thebibliography}
\end{document}